\documentclass[conference]{IEEEtran}
\usepackage{graphicx}
\usepackage{dblfloatfix}    % To enable figures at the bottom of page
\usepackage[compress]{cite}
\usepackage{psfrag}
\usepackage{subfig}
\usepackage{lipsum}
\usepackage{multirow}
\usepackage[cmex10]{amsmath}
\usepackage{amsfonts}
\usepackage{amssymb}
\begin{document}
\title{Testable Design of Repeaterless Low Swing On-Chip Interconnect}

%\author{\IEEEauthorblockN{XXXXXXXXXXXXXXXXXXXXXXXXXXX}
%\IEEEauthorblockA{XXXXXXXXXXXXXXXXXXXXXXXXXXXXXXXX\\
%XXXXXXXXXXXXXXXXXXXXXXXXXXXXX\\
%Email: XXXXXXXXXXXXXXXXXXXX}}

\author{\IEEEauthorblockN{K. Naveen* and Dinesh K. Sharma}
\IEEEauthorblockA{Department of Electrical Engineering, Indian Institute Institute of Technology Bombay,\\
Powai, Mumbai - 400076, India\\
*Email: naveen@ee.iitb.ac.in}}

\maketitle

\begin{abstract}
Repeaterless low swing interconnects use mixed signal circuits to achieve 
high performance at low power. When these interconnects are used in
large scale and high volume digital systems their testability 
becomes very important. This paper discusses 
the testability of low swing repeaterless 
on-chip interconnects with equalization and clock synchronization. 
A capacitively coupled transmitter with a weak
driver is used as the transmitter. The receiver samples the low swing 
input data at the center of the data eye and converts it to 
rail to rail levels and also synchronizes the data
to the receiver's clock domain. The system is a mixed signal circuit
and the digital components are all scan testable. For the analog
section, just a DC test has a fault coverage of 50\% of the structural 
faults. Simple techniques allow integration of the analog components
into the digital scan chain increasing the coverage to 74\%. Finally,
a BIST with low overhead enhances the coverage to 95\% of the structural faults.
The design and simulations have been done in UMC 130 nm CMOS technology.
\end{abstract}


\section{introduction}
Low swing repeaterless interconnects have been researched extensively
for improving the performance of long interconnects, while
keeping the power consumption to acceptable levels 
\cite{Mensink-isscc-2007,RHo-isscc-2007,kim-jssc2010,Lee-jssc14}.
These techniques use low swing on the line with 
equalization to enhance the bandwidth. 
Among the proposed architectures 
for these circuits, the capacitively coupled transmitter is 
one of the most promising
architectures due to its simplicity and robustness 
\cite{RHo-jssc-2008,Mensink-jssc-2010,naveen_vlsi13}.
Such interconnect links
have unknown latencies which can be multiple cycles, thus
requiring appropriate clock synchronizing circuits at the receivers
\cite{Lee-jssc14}. The interconnect design reported in \cite{Lee-jssc14}
uses current mode signaling. At the receiver, a digitally controlled
delay line is 
used to generate a set of clock phases and a foreground calibration
routine selects the phase closest to the center of the data eye. 
Though the system has the advantage of using digital circuits for
clock synchronization,
it has limitation of phase quantization error and it cannot 
track environmental changes without breaking normal operation. The 
authors of \cite{naveen-arxiv15} cite these problems and propose
a background phase synchronizer circuit that uses digital coarse 
correction with analog fine correction.
Thus, repeaterless interconnects need to use sophisticated
mixed signal circuits for achieving high performance and robustness. 
While these repeaterless interconnect solutions have shown
quite some promise, unless these circuits are testable they will
not be appealing for large scale deployment. The digital components of 
these circuits are typically simple and can be tested using fairly 
standardized test methods. However, testing the analog components
along with the digital systems in large designs is a
challenging problem.

This paper discusses the testability of repeaterless low swing 
interconnects that use mixed signal circuits for achieving best performance.
The transmitter uses the capacitively coupled feed-forward equalizer reported in 
\cite{naveen_vlsi13}. A receiver that employs coarse digital correction
and fine analog correction for accurate adaptive synchronization is used
\cite{naveen-arxiv15}.
While this paper discusses testing of low swing interconnect using the above 
circuits as transmitters and receivers, the solutions presented can be used 
for other low swing interconnect systems as well.
The receiver synchronizer circuit
is similar to a phase or delay locked loop. BIST of PLL's is 
generally performed by
adding delays to the inputs of the phase detector and
capturing the divider's outputs \cite{bist-pll-tim05}. Such techniques
are not attractive for interconnect test as it will need
adding delays in the clock or data path. Fault based testing
that does not interfere with the critical path is preferred
for such circuits \cite{bist-pll-tcas2-01}. 
The interconnect test in this paper uses standard
scan test for the digital components. Since the
digital circuits are simple, a 100\% coverage is possible.
For the analog sections, just
a DC test of the full link can detect 50.4\% of the structural faults.
Simple techniques are used to integrate the analog components into
the digital scan chain which enhances the fault coverage to 74.3\%.
The fault coverage can be increased to 94.8\% by using a BIST with a
lock detector. The fault sets covered by the scan test and BIST are
intersecting but not subsets of each other, 
which means to achieve 94.8\% coverage both tests
are required. The circuits do not alter the critical path of the
design.

\subsection{Notations and Fault models}
The additional circuitry added only for the purpose of testing are shaded
grey in all the figures. These circuits will be turned off in normal operation.
The structural fault model \cite{bist-pll-tcas2-01} is used 
for the analog circuits.
\begin{figure*}[t!]
\centering
\psfrag{CkRx}{\small{$\phi_{Rx}$}}
\psfrag{CkTx}{\small{$\phi_{Tx}$}}
\psfrag{Ckd}{\small{$\phi_d$}}
\psfrag{Data}{\small{Data}}
\psfrag{Input}{\small{Input}}
\psfrag{Alex}{\small{Alexander}}
\psfrag{Phase}{\small{Phase}}
\psfrag{Detector}{\small{Detector}}
\psfrag{Charge}{\small{Charge}}
\psfrag{Pump}{\small{Pump}}
\psfrag{(weak)}{\small{(weak)}}
\psfrag{(strong)}{\small{(strong)}}
\psfrag{Logic}{\small{Logic}}
\psfrag{UPDN}{\small{UP DOWN}}
\psfrag{UP}{\footnotesize{$UP$}}
\psfrag{UPst}{\footnotesize{$UP_{st}$}}
\psfrag{DNst}{\footnotesize{$DN_{st}$}}
\psfrag{DN}{\footnotesize{$DN$}}
\psfrag{Counter}{\small{Counter}}
\psfrag{Divider}{\small{Divider}}
\psfrag{Switch Matrix}{\small{Switch Matrix}}
\psfrag{DLL}{\small{DLL}}
\psfrag{VCDL}{\small{VCDL}}
\psfrag{Window}{\small{Window}}
\psfrag{Comparator}{\small{Comparator}}
\psfrag{Retimed Data}{\small{Retimed Data}}
\psfrag{UD}{\footnotesize{$\overline{UP}/DN$}}
\psfrag{EN}{\footnotesize{$Enable$}}
\psfrag{VC}{\small{$V_{c}$}}
\psfrag{Qs}{\footnotesize{$Q_{0-9}$}}
\psfrag{Fine}{\small{Fine tuning loop}}
\psfrag{Coarse}{\small{Coarse tuning loop}}
\psfrag{phin}{\small{$ $}}
\psfrag{CCFFE}{\small{Capacitive}}
\psfrag{Transmitter}{\small{Transmitter}}
\psfrag{SCA}{\small{Scan chain A}}
\psfrag{SCB}{\small{Scan chain B}}
\psfrag{SCK}{\small{Scan clock}}
\psfrag{data}{\small{data}}
\psfrag{Lock}{\small{Lock}}
\psfrag{Detector}{\small{Detector}}
\psfrag{SE}{\small{$S_{en}$}}
\psfrag{TE}{\small{$T_{en}$}}
\psfrag{fsm}{\small{(FSM)}}
\psfrag{test}{\small{$T_{en}$}}
\includegraphics[width=\textwidth]{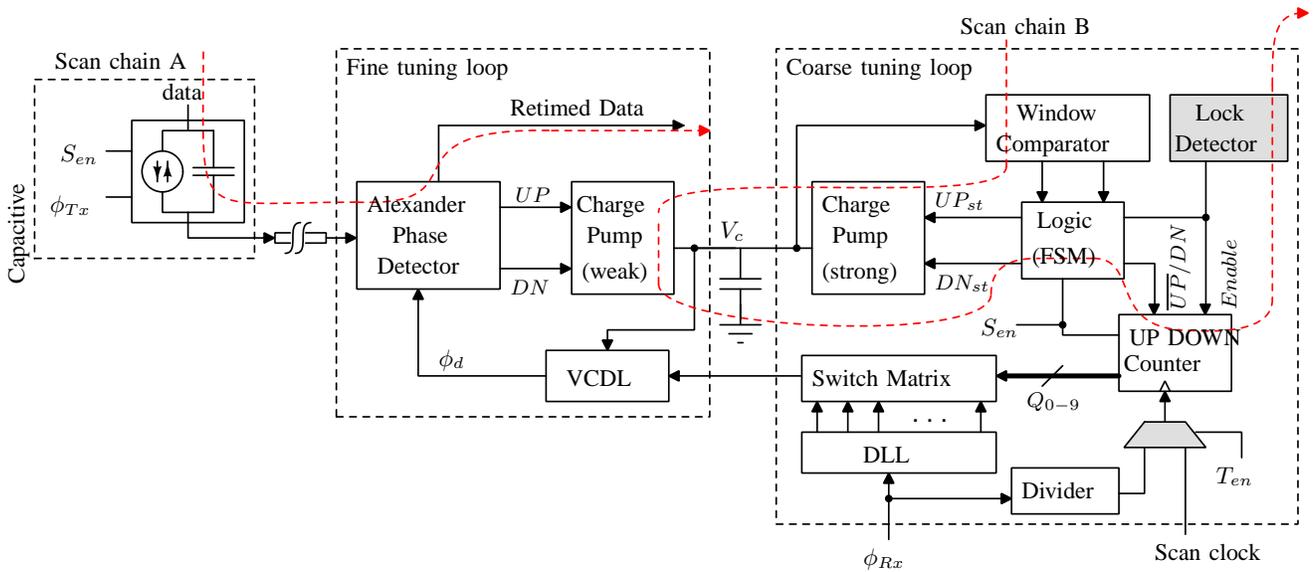}
\caption{Block diagram clock synchronizer system, divided
into fine tuning and coarse tuning loops.
\mbox{VCDL: voltage controlled delay line},
\mbox{$Vc$: control voltage}, \mbox{$S_{en}$: Scan enable signal},
$\phi_{Tx}$: Transmitter clock phase,
$\phi_{Rx}$: Receiver clock phase,
$\phi_d$: sampling clock phase,
$T_{en}$: Test mode enable.}
%\mbox{$T_{en}$ - Toggle enable signal}}
\label{fig:link-scan}
\end{figure*}

\subsection{Paper organization}
The paper is organized as follows. Section II discusses the 
architecture of the capacitively coupled transmitter and the clock 
synchronizer with their test circuits. Section III
describes BIST of the link. Section IV discusses simulation
results and Section V concludes the paper.

\section{Capacitively coupled transmitter and clock synchronizing receiver}
\begin{figure}[h]
\centering
\psfrag{Vc__}{\scriptsize{$V_L$}}
\psfrag{VH__}{\scriptsize{$V_c$}}
\psfrag{VL__}{\scriptsize{$V_H$}}
\psfrag{Timemicros}{\small{Time ($\mu s$)}}
\psfrag{Voltage}{\small{Voltage (V)}}
\psfrag{ph0}{\small{$\phi_0$}}
\psfrag{ph1}{\small{$\phi_1$}}
\psfrag{ph2}{\small{$\phi_2$}}
\psfrag{ph3}{\small{$\phi_3$}}
%\subfloat[Control voltage]{
\includegraphics[width=8cm]{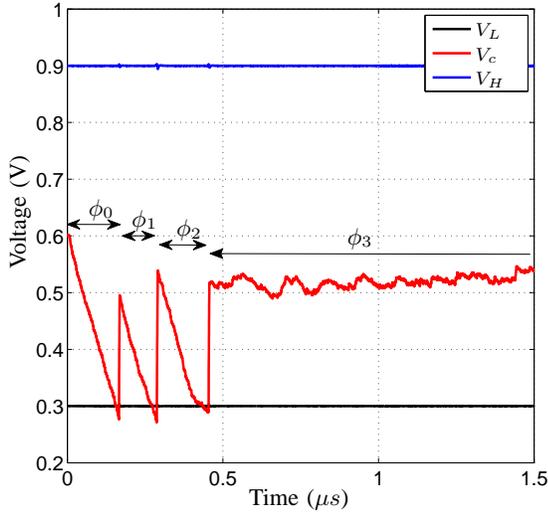}\label{fig:Vc_alex}
%\label{fig:vc}}
%\hspace{5ex}
%
%\subfloat[Ring counter state]{
%\psfrag{Q0_}{\small{$Q_0$}}
%\psfrag{Q1_}{\small{$Q_1$}}
%\psfrag{Q2_}{\small{$Q_2$}}
%\psfrag{Q3_}{\small{$Q_3$}}
%\psfrag{Q4_}{\small{$Q_4$}}
%\psfrag{Q5_}{\small{$Q_5$}}
%\psfrag{Q6_}{\small{$Q_6$}}
%\psfrag{Q7_}{\small{$Q_7$}}
%\psfrag{Q8_}{\small{$Q_8$}}
%\psfrag{Q9_}{\small{$Q_9$}}
%\psfrag{UPDN____}{\small{$\overline{UP}/DN$}}
%\psfrag{Enable__}{\small{$Enable$}}
%\psfrag{UPStrong__}{\small{$UP_{strong}$}}
%\psfrag{DNstrong__}{\small{$DN_{strong}$}}
%\includegraphics[width=5cm]./Qs}
%\label{fig:Qs}}
\caption{Evolution of the fine correction control voltage $V_c$ 
and coarse correction DLL phase of the synchronizer from startup to 
lock condition.}
\label{fig:results}
\end{figure}

Fig. \ref{fig:link-scan} shows the block diagram of the repeaterless interconnect.
The transmitter is the capacitively coupled transmitter from \cite{naveen_vlsi13}.
The receiver in Fig. \ref{fig:link-scan} is a clock recovery circuit
that is used to generate a sampling clock that samples the data at the
center of the data eye \cite{naveen-arxiv15}. 
The circuit has two control loops for coarse and fine phase 
correction. The coarse phase correction loop performs correction in 
discrete steps quantized to the DLL phases. The fine correction loop performs
continuous correction using a voltage controlled delay line.
The circuit uses a phase detector to sense the
phase difference between the received data and the sampling clock. The error
signal from the phase detector is integrated and the integrated output ($V_c$)
controls a VCDL which delays the sampling clock.
The negative feedback loop thus formed, pushes the sampling clock to the 
center of the eye. The VCDL is designed to have a range greater than one
phase step of the DLL over a range of control voltage corresponding to
the window comparator thresholds.
% in Fig. \ref{fig:link-scan}.
If the fine control loop fails to lock while the control voltage
$V_c$ is within the window thresholds, a coarse phase correction request
is issued by the window comparator and the control voltage is reset to 
lie within the window by the strong charge pump. This process repeats
till lock is achieved. In steady state, the DLL phase chosen is within
the VCDL range from the center of the data eye and the control
voltage tunes the VCDL to sample the data at the center of the data eye.
Fig. \ref{fig:results} shows the waveform of the control voltage and 
the chosen phase of the DLL with time,  
as the circuit locks to the correct phase from startup.
The simulated circuit was designed in UMC 130 nm CMOS technology with 
a supply voltage of 1.2 V and a data rate of 2.5 Gbps.

Once lock is achieved,
the phase difference between the sampling clock and the receiver clock
can be found from the coarse tuning control word to an accuracy within the
VCDL phase tuning range. If the sampling clock is less than half cycle
from the receiver's clock, the data is delayed by half a clock cycle 
to ensure reliable crossover to the receiver clock domain.

Since the receiver and 
the transmitter operate in different clock domains, the circuit 
is tested using two separate scan chains
which are the data path scan chain (Scan chain A) and the clock control path
scan chain (Scan chain B). 
The data path scan chain begins at the transmitter, goes through the interconnect
and the phase detector at the receiver. The output of this scan chain is the
 retimed data output of the phase detector.
The clock control path scan chain begins at the window comparator, goes through
the strong and the weak charge pumps, the control FSM, UP DOWN counter and
finally the lock detector block.
When test is enabled, the coarse correction loop's clock input
is driven from an external scan clock as shown in Fig. \ref{fig:link-scan}.
The divider in this circuit can be shared across multiple such receivers in the
chip and tested separately.

%After a short description on the working of the transmitter and 
%the clock synchronizing receiver in the following subsections, the scan test
%of the link will be described in section \ref{sec:scan-test}.

%\subsection{Capacitively coupled transmitter}
%Fig. \ref{fig:FFE_model} shows the capacitively coupled transmitter that 
%uses series capacitors to boost the high frequency content of the data,
%compensating for the low pass characteristics of the interconnect. The 
%two capacitors effectively form a two bit feedforward equalizer. The weak driver
%ennables arbitrarily low data activity factor. The values of the capacitors are 
%optimized using the worst case design method described in \cite{naveen_vlsi13}. 
%A single ended version is shown for the brevity, but actual :0
%implementation used a differential interconnect.

\subsection{The data path scan chain}
\begin{figure}[h!]
\centering
\psfrag{Cs}{$C_s$}
\psfrag{gm}{\small{$-g_m$}}
\psfrag{Csa}{$C_{s\alpha }$}
\psfrag{RL}{$R_L$}
\psfrag{Vcm}{$V_{cm}$}
\psfrag{data}{data}
\psfrag{Scanin}{\small{Scan in}}
\psfrag{Scanout}{\small{Scan out}}
\psfrag{Line}{\footnotesize{Line}}
\includegraphics[width=\columnwidth]{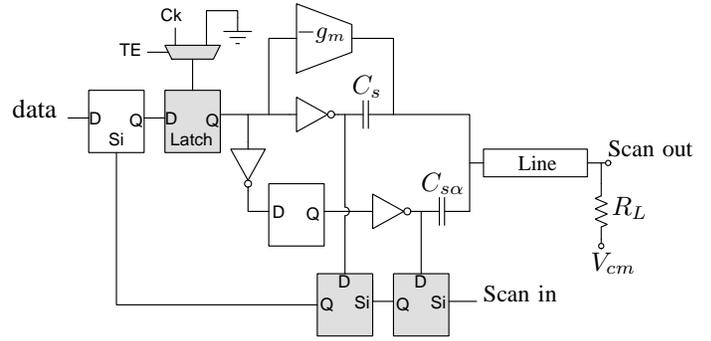}
\caption{The capacitive feed-forward equalizer with the weak driver. 
All the flip flops are clocked by the same transmitter clock, which
is not shown.}
\label{fig:FFE_model}
\end{figure}

The low swing transmitter (Fig. \ref{fig:FFE_model}) is the first component
of the data path scan chain. It uses series 
capacitors to boost the high frequency content of the data, thus
compensating for the low pass characteristics of the interconnect. The 
two capacitors effectively form a two bit feed-forward equalizer. A weak driver
in shunt with the capacitors drives the interconnect with a current source which
enables arbitrarily low data activity factors. The values of the capacitors are 
optimized using the worst case design method described in \cite{naveen_vlsi13}. 
A single ended version is shown for brevity, but actual 
implementation used a differential interconnect. Flip-flops
are added to probe the driver side of the series capacitors (the shaded
flip-flops in Fig. \ref{fig:FFE_model})
and thus enable the scan chain to cover all the nodes up to the 
series capacitors. The additional latch in the data path is added
to optionally introduce a half cycle delay at the transmitter, which 
is required for testing the phase detector at the receiver. This latch 
is transparent during normal operation and it can be absorbed into the 
buffer that drives the line. Since the interconnect has high  
latency, it is not in the critical path. Hence the delay added by the latch
does not degrade the maximum operating frequency of the system.

\begin{figure}[h!]
\centering
\psfrag{Window}{\footnotesize{Window}}
\psfrag{Comparator}{\footnotesize{Comparator}}
\psfrag{offset}{\scriptsize{(with 15mV offset)}}
\psfrag{fig}{\scriptsize{(Fig. \ref{fig:dc-comp}})}
\psfrag{fig2}{\scriptsize{(Fig. \ref{fig:window-comp-100})}}
\psfrag{Rxn}{\small{$R_x^-$}}
\psfrag{Rxp}{\small{$R_x^+$}}
\psfrag{C}{\tiny{C}}
\psfrag{Vm}{\small{$V_{mid}$ (From Fig. \ref{fig:logic})}}
\psfrag{level}{\small{3 level}}
\psfrag{flash}{\scriptsize{Flash ADC}}
\psfrag{Scanin}{\small{Scan in}}
\psfrag{Scanout}{\small{Scan out}}
\includegraphics[width=8cm]{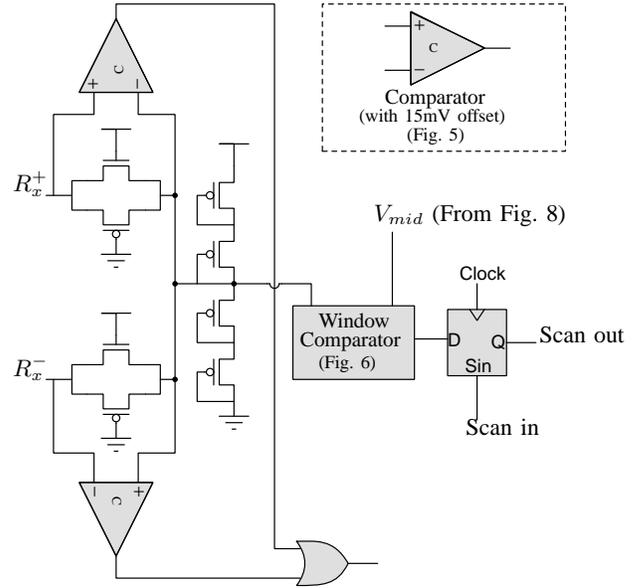}
\caption{The termination of the interconnect at the receiver.}
\label{fig:term-adc}
\end{figure}

\begin{figure}[h]
\centering
\psfrag{+}{\small{$in^+$}}
\psfrag{-}{\small{$in^-$}}
\psfrag{vbn}{\small{$V_{bn}$}}
\psfrag{p2by1}{\small{$\frac{0.8\mu}{0.5\mu}$}}
\includegraphics[width=5cm]{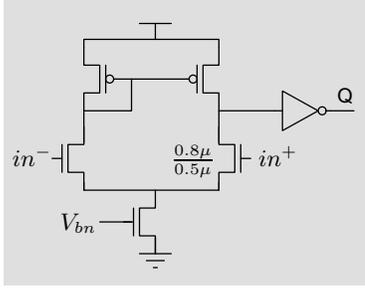}
\caption{Comparator with offset used in the receiver termination. All
un-labelled transistors have W/L=$0.5\mu /0.5\mu$}
\label{fig:dc-comp}
\end{figure}

Fig. \ref{fig:term-adc} shows the circuit
diagram of the receiver termination with the test circuit.
It uses comparators with programmed offset for the DC test. 
This circuit is a single 
stage opamp followed by an inverter (Fig. \ref{fig:dc-comp}). 

The input transistors of the comparator in Fig. \ref{fig:dc-comp} are 
deliberately mismatched to have an offset of 15 mV.
The interconnect is designed for a logic swing of 60 mV and
when the circuit has no faults the comparator gets an input of 30 mV. 
The input transistor sizes are $0.5\mu /0.5\mu $ and $0.8\mu /0.5\mu $,
which is sufficient to overcome any mismatch due to the manufacturing process.
The window comparator compares the bias generated
at the receiver with another voltage divider bias generator in the clock recovery
circuit. The window comparator is constructed using two comparators with a
programmed offset of +15 mV and -15 mV. 
Any fault in the weak driver or the series capacitors at the transmitter
or the termination resistor at the receiver,
results in a mismatch in the two arms of the differential interconnect. 
All such faults are detected by the comparators. Some faults, like drain open
in one of the transistors of the transmission gate resistor, result in a 
dynamic mismatch. This is not detectable at DC. Hence the window comparator
is designed to operate at the scan frequency (which is assumed to be 100 MHz) 
and these faults are detected with a simple toggling data pattern during scan.
Common centroid layout techniques can be used to reduce the inherent 
offset in these comparators. Since these comparators are either used at 
DC or at scan frequencies, the additional parasitics of the common centroid
layout are not a problem.
\begin{figure}[h!]
\centering
\psfrag{p}{\small{$in^+$}}
\psfrag{m}{\small{$in^-$}}
\psfrag{vbn}{\small{$V_{bn}$}}
\psfrag{ck}{\scriptsize{$Clock$}}
\psfrag{8by5}{\small{$\frac{0.8\mu}{0.5\mu}$}}
\includegraphics[width=\columnwidth]{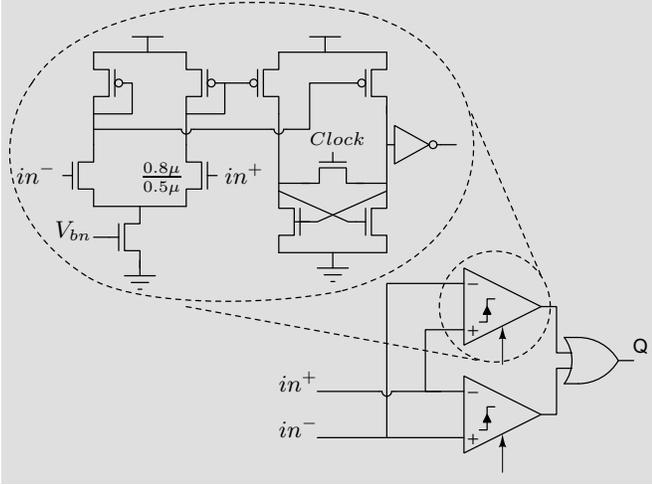}
\caption{Window comparator used in the receiver termination circuit. All
un-labelled transistors have W/L=$0.5\mu /0.5\mu$}
\label{fig:window-comp-100}
\end{figure}

\begin{figure}
\centering
\psfrag{phii}{\small{$\phi _i = \phi_{Rx}\hspace{0.2em} or \hspace{0.2em} \overline{\phi_{Rx}}$}}
\psfrag{chainA}{\small{Scan chain A}}
\psfrag{UP}{\small{$UP$}}
\psfrag{DN}{\small{$DN$}}
\psfrag{Clock}{\small{$Clock$}}
\psfrag{in}{\small{$inp$}}
\includegraphics[width=9cm]{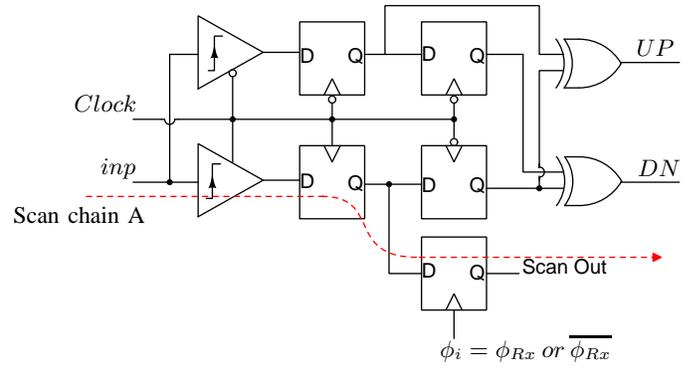}
\caption{The Alexander phase detector with scan.}
\label{fig:apd-scan}
\end{figure}

The last component in the data path scan chain is the Alexander phase
detector at the receiver. The circuit diagram of the phase detector is shown in
Fig. \ref{fig:apd-scan}. When the link is operated at the scan frequency,
the phase detector always asserts the UP signal. To test the other signal path,
the half cycle delay at the transmitter side is enabled, which makes the phase
detector assert the DN signal. Thus, with two passes both these paths
can be tested. The last flip-flop (which is driven
by either $\phi _{Rx}$ or $\overline{\phi _{Rx}}$) in the data path is 
used to insert either a 1 clock or a half clock cycle delay.
This is required for transferring the
data to the receiver clock domain. Half cycle delay is chosen when the 
sampling clock is less than half
a clock cycle away from the receiver clock and this is done
by driving this last flip-flop with $\overline{\phi _{Rx}}$.
For test purposes this can be
controlled via the clock control path scan chain.
When $\phi_{Rx}$ is chosen, it results in an increase in the length of 
Scan chain A register by 1 flip-flop.

\subsection{Clock control path scan chain}

\begin{figure*}[h]
\psfrag{UPS}{\small{$UP_{st}$}}
\psfrag{DNS}{\small{$DN_{st}$}}
\psfrag{UP}{\small{$UP$}}
\psfrag{UPB}{\small{$\overline{UP}$}}
\psfrag{DN}{\small{$DN$}}
\psfrag{DNB}{\small{$\overline{DN}$}}
\psfrag{SEB}{\small{$\overline{S_{en}}$}}
\psfrag{SE}{\small{$S_{en}$}}
\psfrag{Vbp}{\small{$V_{bp}$}}
\psfrag{Vbn}{\small{$V_{bn}$}}
\psfrag{C}{\tiny{C}}
\psfrag{D}{\tiny{D}}
\psfrag{Q}{\tiny{Q}}
\psfrag{RST}{\tiny{RST}}
\psfrag{CP}{\small{CP}}
\psfrag{BIST}{\small{BIST}}
\psfrag{Clock\_gate}{\small{$Enable$}}
\psfrag{Clock}{\small{Clock}}
\psfrag{chainB}{\small{Scan chain B}}
\psfrag{scanin}{\small{Scan in}}
\psfrag{scanout}{\small{Scan out}}
\psfrag{UPDN}{\small{$\overline{UP}/DN$}}
\psfrag{Vc}{\small{$V_c$}}
\psfrag{Vp}{\small{$V_p$}}
\psfrag{0.75}{\small{$V_H$}}
\psfrag{0.25}{\small{$V_L$}}
\psfrag{0.5}{\small{$V_{mid}$}}
\psfrag{200}{\small{$200fF$}}
\centering
\includegraphics[width=14cm]{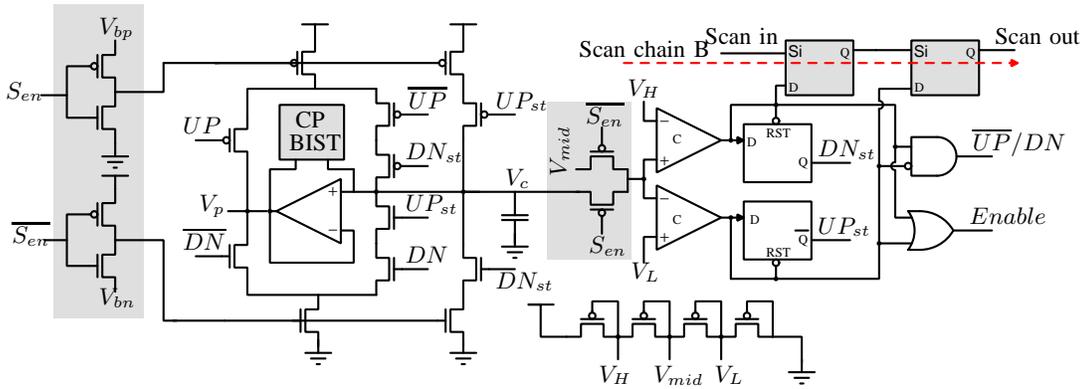}
\caption{Control logic for generating $\overline{UP}/DN$
and $Enable$ signals for the ring counter and 
$UP_{st}$ \& $DN_{st}$ signals
for the strong charge pump. $V_H$, $V_L$ are the
upper and lower thresholds of window comparator respectively. All
flip-flops are clocked with the divided clock of the coarse control
loop.}
\label{fig:logic}
\end{figure*}

%The Scan Chain B is tested at 175 MHz which is the speed at which
%the circuits in the chain operate in the actual system. 
The first
circuit block in this scan chain comprises of the window comparator, the
charge pumps and the control FSM. Fig. \ref{fig:logic} shows the
circuit diagram of this part of the system. The charge pump
is an analog circuit and cannot be included in the scan chain 
as is. To work around this problem the charge pump
is converted to a combinational circuit when scan test is enabled.
This is done by connecting 
the bias voltages for the current sources in the charge pump
to GND for the PMOS source and to VDD for the NMOS sink. 
This
essentially converts the analog charge pump into a combinational
circuit with two inputs UP and DN and one output. Also when scan is
enabled, the window comparator's input is connected to the middle of
the thresholds thus forcing the output to be ``00". 
Two flip-flops are used to capture the comparator's outputs, which
are read through the clock control path scan chain.

Scan chain A is used to make the phase detector assert either UP
or DN signal. This results in the control voltage $V_c$ being driven
to a logic `1' or ``0' respectively. When the scan is disabled and 
the circuit is clocked before re-enabling scan mode, the control FSM
resets the control voltage to within the window. Most of the
faults in the charge pump result in the control voltage not 
being reset to within the window or in not being driven to 
the desired logic level when scan is enabled. 
The comparators outputs can detect most faults in the charge pump circuit.

To test the ring counter, it is first preloaded with a 1 hot value. 
Scan chain A is used to drive the
control voltage in the charge pump to logic `1' or `0'. The 
window comparators outputs enable the ring counter in this
condition and the $UP/\overline{DN}$ signal is driven to `1' 
or `0' depending on the control voltage. This completes the pre-load
step and by de-asserting scan enable ($S_{en}$) and clocking
the circuit the ring counter can be made to count in the desired 
direction before enabling scan again and reading the results.

The last circuit block in the clock control path scan chain is the
switch matrix. Defects in this block may lead to inability of selecting
a desired phase for locking or inability to deselect a particular phase.
This is tested by pre-loading the ring counter with all zeroes pattern. This
causes none of the phases to be picked, resulting in Scan chain A not getting
clocked.
Simple continuity test of Scan Chain A can detect a permanently selected phase.
Further pre-loading different one hot values into the ring counter and testing 
the continuity of
Scan Chain A all the paths in the switch matrix can be tested.

\section{BIST}
The Lock Detector in Fig. \ref{fig:link-scan} is a simple saturating UP counter. 
It logs the number of times a coarse correction request has been issued. 
From any initial 
condition, the number of coarse corrections needed can be no more than half 
the number of DLL phases. The design used a 10 phase DLL and hence a 3 
bit saturating UP counter is sufficient for the lock detector.
For BIST the interconnect is
run with random data at speed. The receiver 
is expected to lock within 2 $\mu s$, which corresponds to 5000 cycles at
2.5 Gbps. Some of the faults in the charge pump, which are not detectable
in the scan test, can be detected using this test. During scan
test the charge pump's current sources were used as switches. This
however masks a drain source short fault in the current source transistors.
The BIST with the lock detector can detect such faults.

\begin{figure}[h!]
\centering
\psfrag{p}{\small{$in^+$}}
\psfrag{m}{\small{$in^-$}}
\psfrag{vbn}{\small{$V_{bn}$}}
\psfrag{1byp2}{\small{$\frac{1\mu}{0.2\mu}$}}
\psfrag{p2by1}{\small{$\frac{0.2\mu}{1\mu}$}}
\psfrag{C}{\scriptsize{C}}
\psfrag{Q}{\small{$Q$}}
\includegraphics[width=8cm]{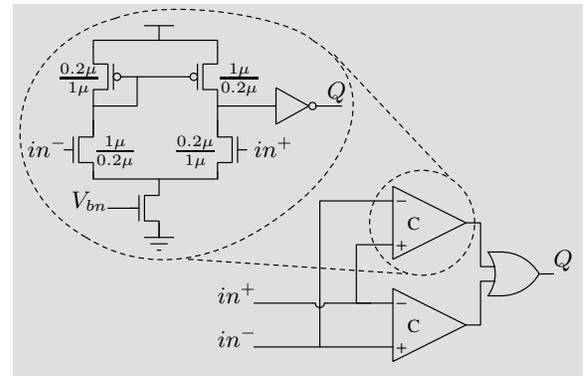}
\caption{The window comparator used in the charge pump for BIST.
All un-labelled transistors have W/L=$0.5\mu /0.5\mu$}
\label{fig:window-comp-dc}
\end{figure}

The scan test of the charge pump tested only the main charge
pump path and the charge balancing path (that drives the node $V_p$ in 
Fig. \ref{fig:logic}) is not tested. Any faults in this second
path or faults in the amplifier in the charge pump, result in the node $V_p$ 
drifting towards $V_{DD}$ or $GND$. 
This pushes one of the current sources to linear
region and as a result causes increased jitter in the recovered clock,
which can degrade the interconnect performance.
The CP-BIST block in Fig. \ref{fig:logic} is a window comparator 
that is designed for a window of 150 mV. Fig. \ref{fig:window-comp-dc} shows the
circuit diagram of the window comparator that is built with two 
comparators with a programmed offset of 150 mV. Once lock is achieved, the 
comparator output being high is an indicator of a fault in the charge pump that
was not detected in the scan test.

The DLL in the receiver is not tested completely by this BIST. This DLL can
be treated as a stand-alone unit and using the techniques 
reported in \cite{bist-dll1,bist-dll2} a complete test of the DLL
can be integrated with the interconnect test.

\section{Simulation Results}
The interconnect system was designed in UMC 130 nm CMOS technology with a 
supply voltage of 1.2 V. The digital components are tested using the scan test.
Since the circuits are logically simple in nature, the stuck at fault
coverage is 100\%. The digital coarse correction is operated at a divided clock 
frequency which is in the range of scan test frequencies. Hence the delay
faults in this path are also tested with 100\% coverage. 

%Architecturally the
%data path scan chain is a simple shift register and delay test with standard 
%at speed scan can cover the delay faults in this path.

\begin{table}[h]
\begin{center}
\caption{Coverage of different types of faults}
\label{tbl:coverage}
\begin{tabular}{|c|c|}
\hline
Defect & Coverage \\ \hline
Gate open & 87.8\% \\ 
Drain open & 93.9\% \\ 
Source open & 93.9\% \\
Gate drain short & 93.9\% \\ 
Gate source short & 100\% \\ 
Drain source short & 100\% \\
Capacitor short & 100\% \\ \hline
Total & 94.8\% \\ \hline
\end{tabular}
\end{center}
\end{table}

\begin{table}[h]
\begin{center}
\caption{Circuit and control input overhead}
\label{tbl:ckt-overhead}
\begin{tabular}{|c|c|}
\hline
Entity & Number \\ \hline
Flip-flop & 7 \\
Comparators (DC) & 4 \\
Comparators (100 MHz) & 2 \\
D-Latch & 1 \\
2$\times$1 Multiplexer & 2 \\
3 bit saturating UP counter & 1 \\
Control signals & 2 \\
Logic gates & 6 \\ \hline
\end{tabular}
\end{center}
\end{table}

For the analog components, two DC tests with the interconnect input at logic 1 and
logic 0 respectively can detect 50.4\% of the structural faults in the circuit.
Scan test of the analog circuits in the receiver by converting the charge pump
to a combinational circuit enhances the coverage to 74.3\%. Most of the faults 
missed by the DC and scan test are detected in the BIST which improves
the fault coverage to 94.8\%. Table \ref{tbl:coverage} tabulates the types 
of faults and their coverage statistics. The total additional circuits required are shown in Table \ref{tbl:ckt-overhead}.

\section{Conclusions}
This paper describes the test of repeaterless low swing interconnects which
use mixed signal circuits. The low swing interconnect considered uses
capacitive feed-forward equalization at the transmitter and 
a clock synchronizing  receiver that uses digital coarse correction 
and analog fine correction. The digital circuits are scan testable easily.
Simple techniques are used to test the analog 
components along with the digital circuits for the scan test. 
A fault coverage of 50\% is achieved with a DC test, which is 
enhanced to 74\% with scan test and to 94\% with a BIST.
This enables the use of low swing interconnect
in large scale high volume digital systems.

\section*{Acknowledgement}
%\vspace{6eM}
The authors would like to thank Prof. Maryam Shojaei Bhaghini and
Prof. Virendra Singh, both from IIT Bombay, for useful discussions. 
The authors would also like to thank Tata Consultancy Services (TCS)
and the SMDP programme of the Government of India for student scholarships
and for providing funds for EDA tools respectively.

\bibliographystyle{IEEE}
\bibliography{IEEEabrv,scan-ref}

\end{document}